\documentclass{IEEEcsmag}

\usepackage[colorlinks,urlcolor=blue,linkcolor=blue,citecolor=blue]{hyperref}

\usepackage{upmath}
\usepackage{cite}

\jmonth{October}
\pubyear{2022}

\begin{document}

\title{Identity, Crimes, and Law Enforcement in the Metaverse}

\author{Hua Xuan Qin}
\affil{Hong Kong University of Science and Technology (Guangzhou)}

\author{Yuyang Wang}
\affil{Hong Kong University of Science and Technology (Guangzhou)}

\author{Pan Hui}
\affil{Hong Kong University of Science and Technology (Guangzhou)\\Hong Kong University of Science and Technology}

\begin{abstract}
With the boom in metaverse-related projects in major areas of the public's life, the safety of users becomes a pressing concern. We believe that an international legal framework should be established to promote collaboration among nations, facilitate crime investigation, and support democratic governance. In this paper, we discuss the legal concerns of identity, crimes that could occur based on incidents in existing virtual worlds, and challenges to unified law enforcement in the metaverse.
\end{abstract}

\maketitle

\chapterinitial{The past years} have seen the realization of various metaverse-related projects in major areas of the public's life\footnote {https://www.forbes.com/sites/michellegreenwald/2022/09/06/5-exploding-areas-of-the-metaverse-that-may-not-yet-be-on-your-radar}, making the metaverse more than just a science-fiction concept in Neal Stephenson's pioneering novel \textit{Snow Crash} or a theoretical framework exclusively for researchers. Given the development strategy of technology giants (e.g. Meta\footnote{https://www.nbcnews.com/tech/tech-news/facebook-goes-meta-zuckerberg-announces-major-restructuring-rcna3605}, Microsoft\footnote{https://www.cnbc.com/2022/01/19/microsoft-activision-what-satya-nadella-has-said-about-the-metaverse.html}, Tencent\footnote{https://www.reuters.com/world/china/tencent-forms-extended-reality-unit-metaverse-race-gathers-steam-sources-2022-06-20/}, and Baidu\footnote{https://www.scmp.com/video/technology/3160931/baidu-unveils-chinas-first-metaverse-platform-xi-rang }) and the expressed interests of large countries (e.g. the United States, China, Japan, South Korea, and the United Arab Emirates) in metaverse-related technologies \cite{Ning}, we might be witnessing only the start of a new era.

\setlength{\parskip}{0pt}

A common vision for the metaverse is a self-sustaining meta-sized virtual space shared by virtual worlds resembling the real world \cite{Lee} without being restricted by time and space \cite{Ning}. The metaverse would not only replicate physical, economic, cultural, and legal elements of the real world but also allow users new ways to interact with them. Users could replicate crimes in the real world in similar or new ways that have a direct impact on the physical world (e.g. theft of property with `real' monetary value) or have no such impact but still seem harmful (e.g. virtual sexual assault). Users could also commit new types of crimes through metaverse-related technologies. One such crime is the simultaneous virtual sexual assault of several users' avatars through a subprogram that forces them to perform sexual acts on each other (graphically or textually - as seen in a case that occurred in LambdaMOO, one of the early text-based virtual worlds\footnote{https://www.villagevoice.com/2005/10/18/a-rape-in-cyberspace/}). All other cases mentioned have also occurred in existing virtual worlds, which can be seen as small-scale prototypes of the metaverse (e.g. Second Life\footnote{https://secondlife.com/} and Horizon Worlds\footnote{https://www.oculus.com/horizon-worlds/}). Since the metaverse will connect similar worlds, it will inevitably inherit their issues.

Because we aim for the metaverse to be unbounded by geographical borders, a single crime could affect several nations, leading to technical and standard-related difficulties during investigation. Allocating all regulation rights to a few would not align with the vision of the metaverse having decentralized, democratic governance \cite{Wang}. To the best of our knowledge, although individual countries have formulated metaverse-related policies\footnote{https://www.hklaw.com/en/insights/publications/2022/09/eu-south-korea-japan-announce-metaverse-regulation-plans}, no universal framework exists yet.

With the metaverse technology boom, we believe that an international legal framework addressing crimes in the metaverse should be established with urgency. This framework would promote inter-nation collaboration, facilitate crime investigation, and support democratic governance. In this paper, we discuss the legal concerns of identity, crimes that could occur, and challenges to unified law enforcement in the metaverse.

We tackle the sections from the law enforcer's perspective. Many works explore ways the developer could protect user privacy, security, and safety. Several works in \cite{Wang}, a September 2022 survey, propose algorithm-based and other technical countermeasures against threats to authentication, data management, privacy, physical safety, and the virtual economy, among others. We will focus on what constraints should be imposed and how to enforce them rather than how developers could implement them.

\section{IDENTITY}

Identity is the set of attributes that makes an individual or a group of individuals distinct from others. From a legal perspective, identity (usually national identity) determines one's rights and duties \cite{Yussef}. In existing virtual worlds, such as Second Life, and the metaverse, users interact with each other and the environment through avatars, digital representations of themselves \cite{Lee}. The legal disconnect between virtual identity and physical-world identity usually complicates law enforcement. In this section, we suggest possible ways to define the legal relationship between the virtual and the physical.

\subsection{Identity of Avatars Controlled by Humans}

For the human user, the avatar can be seen either as an extension of their real-life identity or the manifestation of an entirely different identity, a virtual identity. Part of the attraction of virtual worlds is the veil of anonymity avatars provide, which allows users to experiment without fear of being judged. 

Much debate surrounds the question of whether a user's real-life identity should be exposed when a crime or a misdemeanour is committed. Some believe that anonymity should be preserved depending on the severity. One possible way to achieve this is by treating avatars like companies since they are similar \cite{Cheong}. For instance, they all have to act through a human and cannot be subjected to imprisonment or physical punishment. In this case, avatars become legal persons, who can be sued and represented. Like for companies, the humans behind are not always forced to reveal their identities. Users would need to register their avatars, but unlike for companies in many countries, the real-life identity of the users would not be public.

Another legal concern is whether a user's real-life identity should only be associated with one avatar. The creation of multiple avatars could facilitate defamation \cite{Cheong} or allow a perpetrator to escape punishment under a different avatar (e.g. after they have been banned). If a user can only have one avatar, additional identifying information would be required. Since privacy rights in different countries vary, a consensus on what information users have to provide needs to be reached. If we allow multiple avatars per user, the legal relationship between them needs to be defined. Should all avatars be liable for the crime of one? Should all avatars of a user simultaneously committing the same crime be treated the same way as a single avatar doing so?

Whether an institution can own avatars and have the same rights should also be considered. The answer to the first question in the previous paragraph might vary for companies. For instance, if one of a user's avatars is banned from the metaverse for an offense, banning all his other avatars seems appropriate. However, if one of a company's avatars, which are controlled by different employees, is banned due to the actions of a specific employee, banning all other avatars seems more debatable.

\subsection{Identity of Artificial Intelligence (AI) Avatars}

With advancements in AI and computer graphics, AI avatars resembling human avatars behaviorally and physically might populate the metaverse. They could complete tasks for human users, keep them company, and offer them new ways to experiment with the metaverse. Their presence could however lead to various legal and social issues. For instance, a human user might unknowingly (and unwillingly) become romantically involved with an AI avatar. An AI avatar could also imitate another living or dead person, which can lead to authentication threats and ethical issues. If we take \cite{Cheong}'s suggestion, one way to reduce confusion is to register all avatars - human-controlled and AI-controlled - on a public registry. Each avatar would be registered under a unique registration/ID number. In addition, all AI avatars would be labelled as AI entities. In this case, manufacturers and the avatar registry developers should consider design constraints that enforce this, possibly by adding a required field in the registration form that asks for the avatar's categorization (i.e. human vs AI).

We would also need to consider whether AI avatars should have the same rights and duties as human-controlled avatars. This applies to the AI avatar's role in crimes, to its ownership of digital assets, and to its interactions with others.\\

\noindent\textbf{Crime.} By definition, a crime is an act or a failure to act that results in the harm of one or more victims  \cite [pp. 2-3] {Brenner}. This implies that there is a perpetrator and at least one victim. Can an AI avatar be a perpetrator or a victim? For the former, \cite{Cheong} believes that AI avatars should be indicted like human-controlled avatars under corporate law, but issues could arise. We discuss criminal liability in ``Law Enforcement''.

Perhaps the more important question is whether an AI avatar can be a victim. If not, this nullifies the criminal nature of many acts (i.e. where a human-controlled or an AI avatar harms an AI avatar). Classifying an AI avatar as a victim of permanent damage or destruction seems reasonable. Although the AI avatar acts independently, it is still the creation of a human directly or indirectly (e.g. a human creates an AI avatar that can create other AI avatars). However, should the human receive compensation then? 

Classifying an AI avatar as a victim of crimes that inflict purely emotional damage (e.g. virtual rape) seems less reasonable if we assume that it cannot `feel' like a human. We discuss the implications in ``Criminalization of Fantasy Crimes'' of ``Law Enforcement''.\\

\noindent\textbf{Ownership of digital assets.} Human-controlled avatars would own digital assets - cryptocurrencies (e.g. Bitcoin) or Non-Fungible Tokens (NFTs) \cite{Lee}, which are similar to non-monetary possessions in real life. These assets would satisfy the user's various needs from psychological (e.g. an avatar accessory that shows status and makes the user feel more accepted socially) to physical (e.g. an NFT that can be sold for cryptocurrency, which can be used to buy goods and services for the user's physical survival). An AI avatar has no physical need. If we assume that it cannot `feel', it also has no psychological need. One situation where the ownership of digital assets might be relevant to AI avatars is their learning of prosocial behavior. The gain and loss of such assets could become reward and punishment for the training of the avatar's learning algorithm. More is discussed in ``Criminal Liability of AI Avatars'' of ``Law Enforcement''.\\

\noindent\textbf{Interactions.} We considered the case of a virtual rape of an AI avatar, but this assumes that the AI avatar can engage in sexual acts (e.g. through a script that simulates animation of sex like in Second Life \cite [p. 67] {Brenner}) and could refuse to engage in such acts. Non-consent is a definitive trait of rape and sexual assault \cite [p. 77] {Brenner}. Forbidding developers from giving AI avatars the ability to engage in sexual acts might seem like a direct way to prevent many potential crimes, but this might defeat the purpose of AI avatars for users wishing to experiment without affecting other human-controlled avatars. The same could go for the ability to not consent. Situation-specific constraints for developers might need to be imposed. Such constraints could contain distinctions on the purpose of an AI avatar (e.g. work-oriented and `entertainment'-oriented) and the place where interactions occur (i.e. private or public space). For the former, this information would need to be registered as part of the AI avatar's identity. We should also consider whether an AI avatar could have a marital status. This could lead to new definitions of bigamy.

\subsection{Citizenship and Governance}

In real life, the assignment of legal rights and duties is usually based on nationality. Citizenship is considered the form of nationality with the most privileges\footnote{https://www.britannica.com/topic/citizenship}. Of them, the one most relevant to the metaverse is probably voting right. To ensure democracy, the consensus for governance is a decentralized framework enabled by blockchain and transparent AI algorithms (e.g. decentralized autonomous organizations (DAOs) \cite{Fernandez}) where all necessary (human) members (i.e. users, regulators, developers, and content creators) are involved in the decision making mainly through voting \cite{Wang}. 

It would seem unreasonable to assign voting rights based on the user's real-life citizenship since virtual worlds within the metaverse would be unbounded by space. If governance follows \cite{Fernandez}'s modular-based architecture, where the metaverse is split into modules that can take independent decisions (e.g. for less severe violations) but are still connected to each other, citizenship can be equivalent to the most privileged form of membership in a module. To ensure the quality of the decision, one would need to have been a member of a module for a certain amount of time. A user could have several module-based citizenships, benefit from associated rights, and be subject to associated responsibilities. All users would be metaverse citizens and need to follow the laws from a universal legal framework. Modules can have their own rules as long as these rules do not contradict (offer more freedom than) the metaverse laws. Similar to real-life, these rules could apply to all `residents' (members) or only to citizens of the module.

Citizenship should however not be the only criterion for ensuring vote quality. Generally, real-life countries also consider the age and the criminal record of the citizen. We believe that both could be relevant to the metaverse, should be linked with the user's avatar(s), and should be accessible to the relevant party, e.g. the DAO managing the decision-making process.

There could be a universal minimum legal age. Modules could set higher age restrictions based on their content. A higher legal age would also need to be decided to determine censorship-based rules (for mature content). Such distinction is necessary because certain family-friendly modules could be mainly populated by non-adults. A parallel is the relationship between Teen Second Life \cite [p. 34] {Brenner}, which is Second Life for teenagers (users who have not reached the age to consume mature content), and Second Life, which is for adults and contains mature content. The universal minimum legal age would need to be set to the minimum of the Teen Second Life user's age restriction while restrictions are higher for mature content. Since real-life legal ages vary across the world, a consensus needs to be reached.

Tracking a user's criminal record could be useful in preventing fraud. \cite{Fernandez} suggests a personal reputation-based system that records a user's misbehavior while voting. If we take \cite{Cheong}'s suggestion on making avatars legal persons, such a system would be attached to the avatar instead of the real-life identity of the user. If we allow a user to possess multiple avatars, it would seem reasonable for all avatars to share the reputation system for fraud prevention.

\section{CRIMES}

A crime is a breach of the law, whose purpose is not to obliterate crimes but to contain them to a level where society can function \cite [pp. 2-6] {Brenner}. Traditionally, an act is considered a crime only if it has a direct impact on the physical world - through hard harms (e.g. physical harm on a victim or a property) or certain soft harms, mainly those that harm morality (e.g. gambling, prostitution, and bigamy), that inflict emotional or reputation harm (e.g. stalking and defamation), and that inflict systematic harm (e.g. formation of a monopoly, which discourages market competition) \cite [pp. 6-18] {Brenner}. In the metaverse, like in existing virtual worlds, harm could originate both from the physical and from the virtual, but there is much debate on whether certain acts should be criminalized (punishable by laws) or simply punished locally (by owners of the platforms). 

In this section, we categorize potential crimes in the metaverse as cybercrimes (acts already punishable by law) and fantasy crimes (acts that are usually not criminalized), discuss past or hypothetical situations, and point out legal challenges. We discuss whether fantasy crimes should be criminalized in ``Law Enforcement''.

\subsection{Cybercrimes}

Cybercrimes are generally defined as acts committed through information and communication technologies that violate the law of the physical world \cite{Aigul}. Representative acts are identity theft, online harassment, financial fraud, and cyberextortion crimes\footnote{https://www.fbi.gov/investigate/cyber}. We consider three potentially major cybercrimes.\\

\noindent\textbf{Theft of virtual property with `real' monetary value.} Despite a general agreement on certain crimes, there is a lack of consensus among countries on whether specific acts constitute cybercrimes \cite{Aigul}, especially in virtual worlds, as seen in \cite{Brenner}, which overviews possible virtual world crimes. Countries (and even places within) have different views on whether the same act harms the physical world. One such act that would become increasingly relevant as the physical world's economy blends with that of virtual worlds is virtual property theft.

Many virtual worlds, like Roblox\footnote{https://www.roblox.com/} and Second Life, allow a bidirectional conversion of real currency and their virtual currencies. Virtual items and services (e.g. access to a Roblox game) can be bought with virtual currencies converted from real currency. Users who sold these can then convert the virtual currency they received into real currency. The involvement of real currency has been a motive for stealing virtual property. 

Legal actions have been taken against the theft of virtual property in many countries, such as Japan, South Korea, the Netherlands \cite [pp. 57-58] {Brenner}, and the United Kingdom\footnote{https://publications.parliament.uk/pa/cm201415/cmhansrd/\\cm140721/text/140721w0001.htm\#14072134000055}. The general consensus is that the theft needs to involve the loss of property that has monetary value in the real world. Past cases where legal actions have been taken are related to property directly obtained through real currency\footnote{http://news.bbc.co.uk/2/hi/technology/7094764.stm} or converted into real currency (e.g. through online auction\footnote{https://www.newscientist.com/article/dn7865-computer-characters-mugged-in-virtual-crime-spree/}). However, similar cases have also been disregarded on the basis that the stolen property has no real monetary value - as seen with the cases of Geoff Luurs\footnote{https://www.wired.com/2008/02/police-refuse-t/} and of Qiu Chengwei\footnote{http://news.bbc.co.uk/2/hi/technology/4397159.stm}, the latter escalating into the loss of human life after law enforcement refused to take action.

Proving that a digital property not obtained through real currency (e.g. obtained through currencies derived from virtual jobs, level-based achievements, and events, as seen in many games, or through the digital content creation of a virtual item) has real monetary value can be even more difficult. If the stolen digital property has been subsequently exchanged for real currency (e.g. through an auction), its `real' monetary value could be equivalent to what it was sold for. If not, it seems nearly impossible to determine whether it has `real' monetary value. We discuss this in ``Fantasy Crimes''.\\

\noindent\textbf{Money laundering through the metaverse.} An estimation suggests that 8.6 billion dollars of cryptocurrency has been laundered in 2021. This is a 30\% increase from the previous year\footnote{ https://www.bbc.com/news/technology-60072195}. Such a trend could continue as a decentralized economy (the virtual economy we aim for the metaverse) becomes more accessible since the anonymity it provides makes money funds less traceable.

Some countries have recently taken the initiative in including NFTs as part of their anti-money laundering laws. Specifically, in June 2022, the European Union agreed on anti-money laundering rules\footnote{https://www.reuters.com/technology/eu-backs-crypto-anti-money-laundering-rules-crack-down-dirty-money-2022-06-29/} that require cryptocurrency platforms to report abnormal transactions. These rules also cover transactions exceeding 1000 euros for individuals with unhosted cryptocurrency wallets and service providers. In September 2022, a United States Department of Justice report\footnote{https://aboutblaw.com/43s} demands clarifying NFT platforms' (e.g. auction sites and art galleries) key obligations in combating money laundering since these platforms could be unsure about whether NFTs have real monetary value, a frequent obligation criterion. Also in September, the Japanese government announced its intention in introducing remittance rules for cryptocurrency exchange to track transactions that could be used in money laundering\footnote{https://asia.nikkei.com/Spotlight/Cryptocurrencies/Japan-cryptocurrency-transfer-rules-take-aim-at-money-laundering}. A consensus on laws and potentially platform design constraints would need to be reached. 

One possible design-based way to deter money laundering while not deterring users from using the metaverse due to registration complexity is to require additional identifying information only if a user wants to use their metaverse account for financial transactions. This information would include valid account information from a physical world financial institution (e.g. bank). The user would not need the institution to oversee the transactions (which would contradict the definition of a decentralized economy); the purpose of the account information is to ensure greater credibility and traceability. Such information would also be tied to all of the user's avatars on the registry we have suggested in ``Identity''. For privacy, it could be available only to relevant algorithms (e.g. authentication algorithms).

Additionally, measures could tackle transactions that purely occur in the metaverse (with cryptocurrencies but also with NFTs). More information could be required for transactions exceeding a certain amount. Algorithms could be used to detect abnormal transactions of cryptocurrencies. Two 2022 studies (\cite{Gu} and \cite{Kampers}) respectively propose a three-part framework for anomaly detection in trading volumes - made of on-chain data collection, correlation analysis, and anomaly detection through a long short-term memory deep learning model with an attention mechanism on the output layer - and an unsupervised anomaly detection tool over data streams that combines KDE-Track and isolation-based anomaly detection. One legal concern is whether universal standards on anomaly detection algorithms should be decided.\\

\noindent\textbf{Tax violations in the metaverse.} Common criminal tax violations include tax fraud and tax evasion. Tax violations imply that there is taxation in the first place. Due to varying definitions of cryptocurrency and NFTs, there is little common ground on when cryptocurrency and NFT transactions should be taxed with countries outright banning cryptocurrencies (e.g. China), debating on this (e.g. India), being relatively loose on taxation laws (e.g. Singapore), or collecting capital gains tax (e.g. United Kingdom, Australia, and United States) with varying levels of strictness\footnote{United States: https://coinpedia.org/cryptocurrency-regulation/cryptocurrency-regulation-in-the-usa/\\Other countries: https://www.investopedia.com/cryptocurrency-regulations-around-the-world-5202122}. Tax laws are usually applied based on the related parties' location and citizenship. If our aim is for the metaverse to be unbounded by space, both should not be based on geographical borders. Thus, an international agreement on tax laws would need to be reached for the virtual economy to safely bloom.

Many of the mentioned countries treat cryptocurrencies as property but not legal tender. According to a CNBC article\footnote{https://www.cnbc.com/2021/05/31/cryptocurrency-poses-a-significant-risk-of-tax-evasion.html}, cryptocurrencies facilitate tax evasion because reporting requirements surrounding them are loose. Treating cryptocurrencies as legal tender could be considered for stricter reporting requirements.

Similarly to money laundering, tax violations could be facilitated by the anonymity a decentralized economy provides - specifically by hiding related parties' personal information, such as location and real-life identity. The solutions proposed for money laundering could be applied. On a technical level, a blockchain could enforce tax compliance since its structure makes it difficult to tamper with and supports automatic tax collection. For these reasons, in October 2022, the European Parliament voted in favor of using blockchain technology for taxation\footnote{https://www.europarl.europa.eu/news/en/press-room/20220930IPR41922/meps-call-for-using-blockchain-to-fight-tax-evasion-and-end-crypto-non-taxation}. With precise definitions of cryptocurrencies/NFTs and stricter laws, the blockchain, which is the basis of the metaverse economy, could become a powerful tool in combating tax violation crimes.

One last concern is whether the metaverse should have its own tax system. If it were to have its own law enforcement and public services, this tax system could support these services. If we adopt the modular architecture mentioned in ``Identity'', perhaps there could be universal taxes for the metaverse and module-specific taxes that could help maintain the modules. This would also align with our definition of citizenship in the metaverse (in ``Identity''). 

\subsection{Fantasy Crimes}

Fantasy crimes are acts that would be crimes if committed in the physical world but are usually not when done in the virtual world \cite [p. 62] {Brenner}. We consider potentially major ones.\\

\noindent\textbf{Gambling.} Laws have criminalized gambling with virtual currencies bought with or that can be exchanged for real money to protect the user \cite [p. 54] {Brenner} or to prevent money laundering \cite [p. 59] {Brenner}. In the context of fantasy crimes, we consider gambling with virtual currencies not bought with real money. Generally, there is no legal restriction on it. However, while the user cannot gamble away their real money, addiction could keep them from productivity (e.g. they spend their entire day gambling) or from a healthy lifestyle. Different places across the world have different views on online gambling\footnote{https://worldfinancialreview.com/online-gambling-laws-throughout-world/}. If gambling as a fantasy crime is criminalized, a consensus needs to be reached.\\

\noindent\textbf{Prostitution.} We consider prostitution purely between the adult-looking avatars of real consenting adults. While real-life prostitution is restricted in many countries in varying degrees\footnote{https://globalnews.ca/news/128029/at-a-glance-prostitution-laws/}, it is allowed in Second Life (as seen in this currently existing location\footnote{https://world.secondlife.com/place/bf0deacb-7255-4efd-4706-f2c2867a8ba3}). Since virtual prostitution is done between two virtual adult-looking entities, it has not raised enough controversy to make the site ban it (in contrast to sexual ageplay, as seen later). In Second Life, users engage in prostitution to earn virtual currency that can be converted to real currency\footnote{https://freemmorpg.top/second-life-for-prostitution-a-blogger-spoke-about-the-work-of-a-virtual-brothel-in-second-life/} and/or to fulfil some fantasy\footnote{https://www.wired.com/2009/01/italian-woman-e/}. While virtual prostitution has no risk to the physical health and safety of the humans behind the avatars, if virtual currency becomes as important as real currency, one could consider whether there should be universal financial restrictions.\\

\noindent\textbf{Bigamy.} Currently, in games and virtual worlds, users in monogamist countries can simultaneously be married to several partners regardless of whether they are married in real life because in-game (or virtual) marriages lack legal weight. If the metaverse takes over our life and the economy, virtual marriages could have as many moral and financial (e.g. taxation) implications as real marriages.\\

\noindent\textbf{Theft and other forms of property deprivation.} We consider theft and damage of virtual property (e.g. vandalism) not obtained from real currency and, for theft, not exchanged for real currency. Currently, such acts do not seem to impact the physical world and thus harm social order, but as the virtual economy grows in importance, we consider the impact in ``Law Enforcement''.\\

\noindent\textbf{Virtual murder and physical assault.} Due to the virtual nature of an avatar, acts that can physically harm humans in real life cannot deal any permanent damage unless such a feature has been programmed. An avatar can instead only be damaged through computer-related methods (e.g. hacking). Depending on the avatar's real monetary value, this can be categorized either as a cybercrime or property deprivation as fantasy crime. We consider virtual acts that would harm humans in real life. In existing virtual worlds, they are usually features enabled by users or only available in areas that users enter of their own volition (e.g. battle games in Roblox and areas in Second Life). They are essentially parts of violent video games, which are legal in most countries with some rarely enforced age restrictions (e.g. age ratings). If the same design constraints of existing virtual worlds are applied to the metaverse, the reasoning behind existing laws could be applied. One concern is whether virtual murder or any other form of physical assault enabled through hacking (i.e. non-consensual acts) should be treated simply as hacking given the moral implications.\\

\noindent\textbf{Virtual rape and other non-consensual sexual acts.} Such acts have been committed in both older and new virtual worlds (e.g. Second Life \cite [p. 76] {Brenner}, Roblox\footnote{https://www.bbc.com/news/technology-44697788}, and Horizon Worlds \footnote{https://www.usatoday.com/story/tech/2022/01/31/woman-allegedly-groped-metaverse/9278578002/}). These acts are usually the suggestive touching of another user's avatar, sexual acts on avatars accompanied by suggestive verbal comments, and sometimes\cite [pp. 75-76] {Brenner}, the hacking of the platform to make human-controlled characters perform sexual acts on each other. None can physically harm the victim. Victims also feel varying degrees of emotional distress, some feeling disconnected while others, traumatized \cite [pp. 76] {Brenner}. However, immersive experience supported by new metaverse-related technologies (e.g. virtual reality) could increase one's connection with their avatar and thus emotional harm (as one victim has noted\footnote{https://www.thecut.com/2016/10/woman-sexually-assaulted-by-player-in-quivr-using-htc-vive.html}). Our search reveals sanctions by only owners of the platforms. As a legal expert has stated in the article on the Horizon Worlds case, which occurred in December 2021, it is currently unlikely for legal actions to be taken. We propose an approach in considering the criminalization of such acts in ``Law Enforcement''.\\

\noindent\textbf{Sexual ageplay and child pornography.} In some virtual worlds, a user can make their avatar look different than what those of their chronological age usually look like. This has led to controversies surrounding virtual child pornography, entirely computer-generated sexually explicit graphics of children. A notorious example was Second Life's Wonderland, where adult users engaged in sexual ageplay \cite [p. 91] {Brenner} (i.e. sexual conduct with avatars looking like children - now disallowed in Second Life\footnote{https://wiki.secondlife.com/wiki/Linden\_Lab\_Official:\\Clarification\_of\_policy\_disallowing\_ageplay}). Virtual sexual ageplay is often believed to be part of virtual child pornography. Opinions about this among users were divided. On one hand, some users believed that such an act should not be criminalized because there is no real child involved and there is no apparent causality between virtual child pornography and behavior. This reasoning is in accordance with the laws of certain countries, such as the United States\footnote{https://web.archive.org/web/20220202060126/https://www.\\mtsu.edu/first-amendment/article/4/ashcroft-v-free-speech-coalition}. Some even believed that allowing virtual child pornography could satisfy the desires of potential pedophiles, preventing them from harming real children. On the other hand, others believed that allowing virtual child pornography could encourage violence against real children. Many countries (e.g. Canada\footnote{https://laws-lois.justice.gc.ca/eng/acts/c-46/section-163.1.html} and Australia\footnote{https://www.mondaq.com/australia/crime/895042/books-cartoons-and-dolls-can-amount-to-child-pornography describes legal actions taken against virtual child pornography in cartoons (McEwen v Simmons), books (Traynor v McCullough), and chatroom communications (R v Jarrold).}) also prohibit virtual child pornography to a certain degree. A consensus needs to be reached on whether to criminalize virtual child pornography in the metaverse.\\

\noindent\textbf{Simulation of the Holocaust and sensitive historical events.} \cite [pp. 95-96] {Brenner} mentions the hypothetical situation of the reconstruction of a Nazi death camp where users can roleplay as Nazis and inmates. While this would be illegal in many European countries, it would not be in the United States. For countries to reach a consensus, we might need to consider whether non-educational simulations of sensitive historical events (e.g. genocides, assassinations, and acts of terrorism) should be allowed in general.

\section{LAW ENFORCEMENT}

As seen with cybercrimes, it has been difficult to establish an international legal framework due to countries' varying beliefs on rights and duties \cite{Aigul}. In this section, we propose possible approaches in considering different challenges to the creation of a unified legal framework.

\subsection{Criminalization of Fantasy Crimes}

The traditional definition of a crime is grounded in the physical world because it is the only world we have been living in until the past decades. Even now, virtual worlds remain on the margin of many people's life. The estimated global Internet penetration in 2022 is 63\%\footnote{https://www.statista.com/statistics/269329/penetration-rate-of-the-internet-by-region/}. Virtual worlds are only small parts of the Internet. Thus, it is reasonable to assume that an act purely occurring the virtual would not disturb social order.

However, with the metaverse, our society would become a barely distinguishable blend of the physical and the virtual. Acts purely situated in the virtual world could disturb social order as much as acts that directly impact the physical world. We should thus consider fantasy crimes not in our current context but in the futuristic context of the metaverse.

To illustrate this, we consider a scenario that seems to have no impact on the physical world's order currently: AI avatars cannot be victims of sexual and non-sexual violent acts that do not threaten their existence permanently. If we assume AI avatars cannot `feel', for a fantasy crime that only inflicts emotional harm (e.g. virtual rape or other forms of assault), it seems unreasonable to consider the AI avatar a victim (and thus to consider the act as a crime).

We believe that, while the AI avatar might not need the rights of a victim, not convicting the perpetrator could disturb social order in the long term in a metaverse world. The main purpose of punishment is deterrence. Judges of existing fantasy crime cases (e.g. virtual child pornography \cite[pp. 92-94]{Brenner}) have considered not only the direct physical harm but also whether tolerating the act could lead to the perpetrator adopting a less acceptable behavior (e.g. whether exposure to virtual child pornography can make someone act on real children eventually). We believe such a perspective can also be used for our case.

However, if we base our reasoning on existing research, our answer would be inconclusive. Recent research on exposure to sexual \cite{Malamuth} and non-sexual \cite{Drummond} media content suggests that exposure to media violence (comparable to violence in virtual worlds) could at best increase the risk of aggression for a specific group of people. So, by the logic of current courts, the mentioned fantasy crimes against AI avatars should not be criminalized.

However, existing studies have participants who have mainly grown up in the physical world, where such behavior is usually discouraged. Although theories on the process of moral development might vary (e.g. Jean Piaget's Theory of Moral Judgment focused on stages \cite{Piaget} and the Social Domain Theory being on more gradual development \cite{Smetana}), there is a consensus that a person's environment influences their concept of morality and thus their ethical behavior. The laws and social norms of the real world could have contributed to existing media consumers' appreciation of the (lack of) morality behind violent acts.

Now consider a world where life has blended with the metaverse. Users might be spending most of their time in this place from a very young age. Environmental influence on their concept of morality would mainly come from the metaverse. If violent acts against AI avatars are allowed, the social norms and interactions in that metaverse could lead to the child having a different concept of morality and behaving less ethically in and outside virtual worlds by our society's standards. These standards, although not perfect, are the results of history and ensure a certain equilibrium. If they are broken, there is no telling what the world would become.

Our reasoning is purely theoretical. Studies would need to be done. However, we hope that this highlights the importance of determining the severity of virtual acts of violence. We chose this scenario of AI avatars, which seems to have no relationship with the physical world, to illustrate that, even in this case, social order could be disturbed. 

This reasoning could also be applied to fantasy crimes concerning human-controlled avatars - non-consensual acts (e.g. rape) and certain consensual acts (e.g. sexual ageplay). We do not mean that all these acts should be criminalized or treated the same way as their real-life counterparts if they are criminalized; we are simply providing an approach to evaluate fantasy crimes. If some are criminalized, we suggest specifying the context to ensure that users still benefit from the freedom of experimentation only the metaverse could provide (e.g. private vs public areas). We also believe that punishment of fantasy crimes should be lighter to put more weight on crimes having a direct impact on the physical world since they are often irreversible and more severe (e.g. murder in the real world is always irreversible and rape in the real world can leave permanent physical wounds in addition to emotional damage).

\subsection{Property Deprivation Fantasy Crimes}

Similarly to real life, content creation not directly involving real currency (e.g. certain artistic creations and computer applications) could become a major source of income. Tolerating theft and property deprivation that do not involve real currency could harm the economy. If they are criminalized, appropriate punishment and compensation would need to be decided. For theft, the tracing and return of the stolen property could be prioritized. If this is impossible (e.g. the property has been destroyed, consumed, or permanently damaged), the perpetrator would need to compensate the victim. The theft or damage of avatars whose 'real' monetary value cannot be demonstrated could be treated similarly.

\subsection{Criminal Liability of AI Avatars}

The traditional belief is to punish the manufacturer for an AI entity's actions, but this seems gradually unfairer as AI programs become more complex and unpredictable. Treating AI avatars as legal persons could facilitate certain penalties, such as imprisonment in a virtual jail, which is equivalent to a ban \cite{Cheong}. 

However, this will not address crimes that require financial compensation. Consider the following scenario. An AI avatar `learns' on its own how to corrupt digital assets. It then corrupts a human user's virtual property that could be sold for real currency. If the offending avatar is controlled by a human, they might need to pay some financial compensation. However, if the AI avatar cannot own any assets, it cannot pay. We believe that, similar to cases concerning AI machines \cite{Glavani}, if the incident is caused by a defect in design (e.g. bug), the manufacturer should be held responsible. If there is no defect, as mentioned, this would be unfair. Laws regarding AI avatars in this situation should be discussed. One solution could be the creation of insurance plans for AI-related incidents in the metaverse.

A separate reputation system could also be established, or as mentioned in the ``Identity'' section, the AI avatar could be allowed to own digital assets, which could contribute to its learning of prosocial behavior through reward and punishment. In these cases, what constitutes prosocial and antisocial behaviors need to be established. Loopholes need to be considered to avoid metaverse versions of situations in science fiction writer Isaac Asimov's works on AI robots. One concern would be on preventing AI avatars from being manipulated into committing crimes.

\subsection{Countermeasures to Governance and Digital Forensics Threats}

Several technologies could counter threats to governance and digital forensics on a technical level. One main concern is whether there should be universal standards for the design and use of such technologies. We describe some key technologies covered by \cite{Wang} in the following sections and possible challenges.\\

\noindent\textbf{Governance.} Given the scale of the metaverse, AI algorithms could be used to improve the efficiency of governance. A key use of AI in the metaverse is for detecting misbehaving social accounts or Sybil accounts, which are duplicate accounts created by a single user trying to take over a network. Current research \cite{He} has focused on using a long short-term memory model, an artificial neural network that has the advantage of being able to process sequences of data instead of only single data points, to dynamically reveal malicious accounts' suspicious signals in an online social platform by analyzing textual information generated by users and user activities. Experiments support the detection accuracy of this method for real-life social media accounts, but more would need to be done for metaverse accounts since the avatars they are associated with would generate more complex and abundant data. Current AI algorithms can also be biased (e.g. ethnically), which could lead to major ethical concerns in a multicultural space like the metaverse. Research on AI governance from a sociological perspective \cite{Gasser} proposes a multi-layered conceptual framework that includes, from bottom to top, a technical, an ethical, and a social and legal layers, but more research on the implementation of such structure would need to be done. Specifically, legal and sociological insights would need to be combined with technological knowledge to define what is a malicious account/avatar or a malicious behavior.

Other technologies mainly tackle threats to decentralized governance. Blockchain technologies could be used to reduce centralized governance risks since smart contracts could ensure automated decentralized governance. Existing research \cite{Bai} has proposed combining blockchain with a Stackelberg game approach, which could encourage user participation in administrative processes. Such a method has time efficiency and user utility advantages but possible scalability and security issues when used in real life \cite{Wang}.

Existing research has also focused on countering opportunistic attacks for price manipulation that leads to illicit financial gains with a Dirichlet-based probabilistic model that detects compromised local agents in decentralized systems by evaluating reputation levels using historical operation observations\cite{Li}. Such a detection method has the advantage of being fast but still needs to be tested in a real-world environment to improve efficiency and accuracy.\\

\noindent\textbf{Digital forensics.} Digital forensics is the virtual reconstruction of crimes committed, which supports their investigation, especially during disputes about accountability. By definition, the metaverse is a space connecting a variety of virtual worlds. Due to the high interoperability and variation of these worlds, forensics investigation could be challenging \cite{Wang}. To facilitate the task, different virtual world owners might need to cooperate to reach a consensus on design standards of different platforms.

More research also needs to be done on existing digital forensics technologies. A main way to investigate crimes in the metaverse is through images and videos \cite{Wang}. Prior work \cite{Swaminathan} has focused on developing a general digital camera image forensic mechanism to validate image authenticity (i.e. whether an image is from a specific device) based on the knowledge that in-camera and post-camera image processing leaves unique fingerprints on the image. This method is highly efficient for non-intrusive image forensics but lacks anti-forensics defense. Subsequent research \cite{Stamm} addresses this limitation with a video frame forensics method that has anti-forensics detection. Like other digital forensics methods in the metaverse, this could lead to trust issues and high labor cost. \cite{Wang} proposes to use blockchain to establish trust among different parties and enhance automation during cross-platform digital forensics investigation. Existing research \cite{li2021toward} focuses on employing blockchain for a decentralized forensics method with smart contracts that ensure forensics execution can be audited. According to \cite{Wang}, more research on the benefits of smart contracts in automation needs to be done. Existing research \cite{zou2018multigranularity} also focuses on privacy leakage forensics, which can be used to determine privacy violation accountability, using cloud forensics. This method has high detection efficiency on real samples but is currently limited to specific privacy leakage cases. More research in this area would need to be done.

\section{CONCLUSION}

Laws shape the society by enforcing what is acceptable and what is not. Over time, they become the basis for social norms, moral reasoning, and ethical behavior. When conceptualizing a legal framework, we should not only consider its impact in our current setting but also in the futuristic context where the metaverse has become our life. This might be difficult given the lack of empirical evidence. Fortunately, past and viable situations in existing virtual worlds, theories in psychology, and the reasoning behind existing laws can give us a general idea.

This paper proposes some approaches to elaborating a universal legal framework for the metaverse by defining the legal role of avatars, used by humans and possibly AIs to interact with the metaverse, and the relationship between governance and identity, giving an overview of possible crimes, and describing law enforcement challenges. To ensure the harmony of a world unbounded by space like the metaverse, countries would need to reach a consensus on the rights and duties of the different parties involved.

\bibliographystyle{IEEEtran}
\bibliography{references}

\newpage
\begin{IEEEbiography}{Hua Xuan Qin}{\,}is working on her Ph.D. of Computational Media and Arts at the Hong Kong University of Science and Technology, China. She received a Bachelor for Honours Computer Science with a Psychology Minor from the University of Waterloo, Canada, in 2022. Her research interests surround the metaverse. Contact her at hxqin682@connect.hkust-gz.edu.cn.
\end{IEEEbiography}

\begin{IEEEbiography}{Yuyang Wang}{\,} is a postdoctoral fellow. He received a PhD in computer science from the Arts et Metiers Institute of Technology in France. He worked as a visiting researcher at the Karlsruhe Institute of Technology in Germany. In addition, he holds a master's degree in numerical methods in engineering from the International Center for Numerical Methods, the Polytechnic University of Catalonia in Spain. His research interests include data-driven modelling methods to develop intelligent navigation techniques in virtual environments and create educational tools on the XR platform. Contact him at yuyangwang@ust.hk.
\end{IEEEbiography}

\begin{IEEEbiography}{Pan Hui}{\,} is a Chair Professor of Computational Media and Arts and Director of the Centre for Metaverse and Computational Creativity at the Hong Kong University of Science and Technology (Guangzhou), and a Chair Professor of Emerging Interdisciplinary Areas at the Hong Kong University of Science and Technology. He is also the Nokia Chair in Data Science at the University of Helsinki. Professor Hui received his PhD from the Computer Laboratory at University of Cambridge, and both his Bachelor and MPhil degrees from the University of Hong Kong. He is an International Fellow of the Royal Academy of Engineering, an IEEE Fellow, a member of Academia Europaea, and an ACM Distinguished Scientist. Contact him at panhui@ust.hk.
\end{IEEEbiography}

\end{document}